\documentclass[aps,pra,floatfix,twocolumn,superscriptaddress]{revtex4-1}
\usepackage{amsmath,amssymb}
\usepackage{graphicx}
\usepackage{epsfig}
\usepackage{physics}
\usepackage[normalem]{ulem}
\usepackage[usenames]{color}

\begin{document}

\title{Exciton-Polariton Topological Insulator with an Array of Magnetic Dots}





\author{M.~Sun}
\affiliation{Center for Theoretical Physics of Complex Systems, Institute for Basic Science (IBS), Daejeon 34126, Korea}
\affiliation{Basic Science Program, Korea University of Science and Technology (UST), Daejeon 34113, Korea}

\author{D.~Ko}
\affiliation{Center for Theoretical Physics of Complex Systems, Institute for Basic Science (IBS), Daejeon 34126, Korea}
\affiliation{Basic Science Program, Korea University of Science and Technology (UST), Daejeon 34113, Korea}

\author{D.~Leykam}
\affiliation{Center for Theoretical Physics of Complex Systems, Institute for Basic Science (IBS), Daejeon 34126, Korea}
\affiliation{Basic Science Program, Korea University of Science and Technology (UST), Daejeon 34113, Korea}

\author{V.~M.~Kovalev}
\affiliation{A.V.~Rzhanov Institute of Semiconductor Physics, Siberian Branch of Russian Academy of Sciences, Novosibirsk 630090, Russia}
\affiliation{Department of Applied and Theoretical Physics, Novosibirsk State Technical University, Novosibirsk 630073, Russia}

\author{I.~G.~Savenko}
\affiliation{Center for Theoretical Physics of Complex Systems, Institute for Basic Science (IBS), Daejeon 34126, Korea}
\affiliation{A.V.~Rzhanov Institute of Semiconductor Physics, Siberian Branch of Russian Academy of Sciences, Novosibirsk 630090, Russia}


\date{\today}

\begin{abstract}
Recently there have been several proposals on exciton-polariton topological insulators, most requiring strong external magnetic fields induced by bulky superconducting coils.
We propose an alternate design for a polariton topological insulator, where excitons are in a proximity of an additional layer of a ferromagnetic material with a predefined magnetic moment located between the pillars of the cavity. 
Our design supports a variety of topological phases and transitions between Chern numbers 
$\pm 2$ and $\pm 1$ by varying either the magnetic moment of the ferromagnetic material or the spin-orbit coupling between different spin projections, thus enabling compact polariton devices harnessing switchable topological edge modes. 

\end{abstract}


\maketitle

\section{Introduction}
Robust transport of particles in topologically nontrivial systems occurs due to the existence of protected edge modes at interfaces between media with different topological properties~\cite{Hasan:2010aa,Qi:2011hb,Chiu:2016aa}.
Such edge states are robust against disorder, which makes them appealing from fundamental and application-oriented viewpoints. Phase transitions to topological insulating phases were originally discovered in the context of the integer quantum Hall effect~\cite{Klitzing:1980aa}, where a nonzero quantized Chern number is associated with disorder-robust quantization of the particle current (and conductivity) in strong external magnetic fields. 
Later this idea was extended to the anomalous quantum Hall and quantum spin Hall effects~\cite{Haldane:1988aa, Kane:2005aa, Bernevig:2006aa, Konig:2007aa}, where an external net magnetic field is not required, instead, it is replaced by other effects such as time-reversal symmetry breaking induced by internal magnetization of the material~\cite{Karplus:1954aa, Chang:2013aa} or strong spin-orbit coupling. Topological phases have now been extended to various physical systems, including photonics, cold atoms, and acoustics~\cite{Wang2009,Jotzu:2014aa, Yang:2015aa}.

A recent experiment has demonstrated an exciton-polariton topological Chern insulator~\cite{Klembt:2018aa}. Distinct from other platforms, the nontrivial topological phase arose from a combination of \emph{three} main ingredients: an artificial lattice made of micrometer-sized pillars, excitonic spin-orbit coupling or photonic TE-TM splitting, and an external magnetic field in the Faraday configuration~\cite{Bardyn:2015aa, Nalitov:2015aa}. The lattice creates a finite Brillouin zone with energy bands in reciprocal space, the spin-orbit coupling creates a momentum-dependent energy splitting between spin states without opening a band gap, and the external magnetic field induces a Zeeman splitting between spin-up and spin-down exciton-polaritons. Notably, all three effects are required to create a topological band gap in which the protected edge states reside, in contrast to other platforms, where a lattice combined with a uniform or staggered magnetic field is sufficient. 


So far, theoretical and experimental studies on the exciton-polariton topological insulators have focused on the case of a uniform external magnetic field~\cite{Gulevich:2016aa,Nalitov:2015aa,Bardyn:2015aa,Yi:2016aa,PhysRevB.93.085438}.
In order to observe a measurable bandgap a large external magnetic field (up to 5~T) was used, requiring superconducting coils and cryogenic temperatures~\cite{Klembt:2018aa}.
Evidently, this strong external magnetic field a significant drawback limiting potential applications in real devices, such as using the edge states for the disorder-robust transfer of optical signals and information processing.
In particular, such essential advantages of microcavity samples as their compactness and potential for room-temperature operation are compromised or even negated.
It is therefore crucial to explore application-friendly mechanisms and techniques by which topological polaritons can be created without an external magnetic field.

The existing proposals have been based on such effects as the nonlinear dynamics under resonant pumping~\cite{Mandal:2019aa}, time or spin-dependent pumping~\cite{Sigurdsson:2019aa}, and spontaneous symmetry breaking~\cite{Karzig:2015aa,Janot:2016aa}.
However, all these proposals have their technological limitations. 
For example, using the nonlinear terms due to particle interaction requires strong resonant pumping of the sample in order to excite the edge as a perturbation~\cite{Bardyn:2015aa,PhysRevB.93.085438}; 
the spin-dependent pumping demands independent treatment of each lattice site, which makes the design of even a single experiment tricky, not to say that it makes the creation of nontrivial topology-based logic impossible.   

In this article, we propose an easily realizable alternate design for an exciton-polariton topological insulator in which the external uniform magnetic field is replaced by an internal inhomogeneous magnetization, produced by a magnetic material (MM) embedded within the microcavity~\cite{Otani:1993aa,Vavassori:2000aa}. 
We show that the resulting staggered magnetic field is capable of opening a topological band gap, and can yield topological transitions between different Chern numbers ($\pm 2$ to $\mp 1$) as either the magnetic field or spin-orbit coupling strength are varied, similar to the uniform external field case~\cite{Bleu:2017aa}. 
With our approach, superconducting coils and other bulky equipment can be eliminated, which is an important step towards translating exciton-polariton topological insulators towards actual devices.



\section{System schematic}
\begin{figure}[ht]
	\includegraphics[width=.49\textwidth]{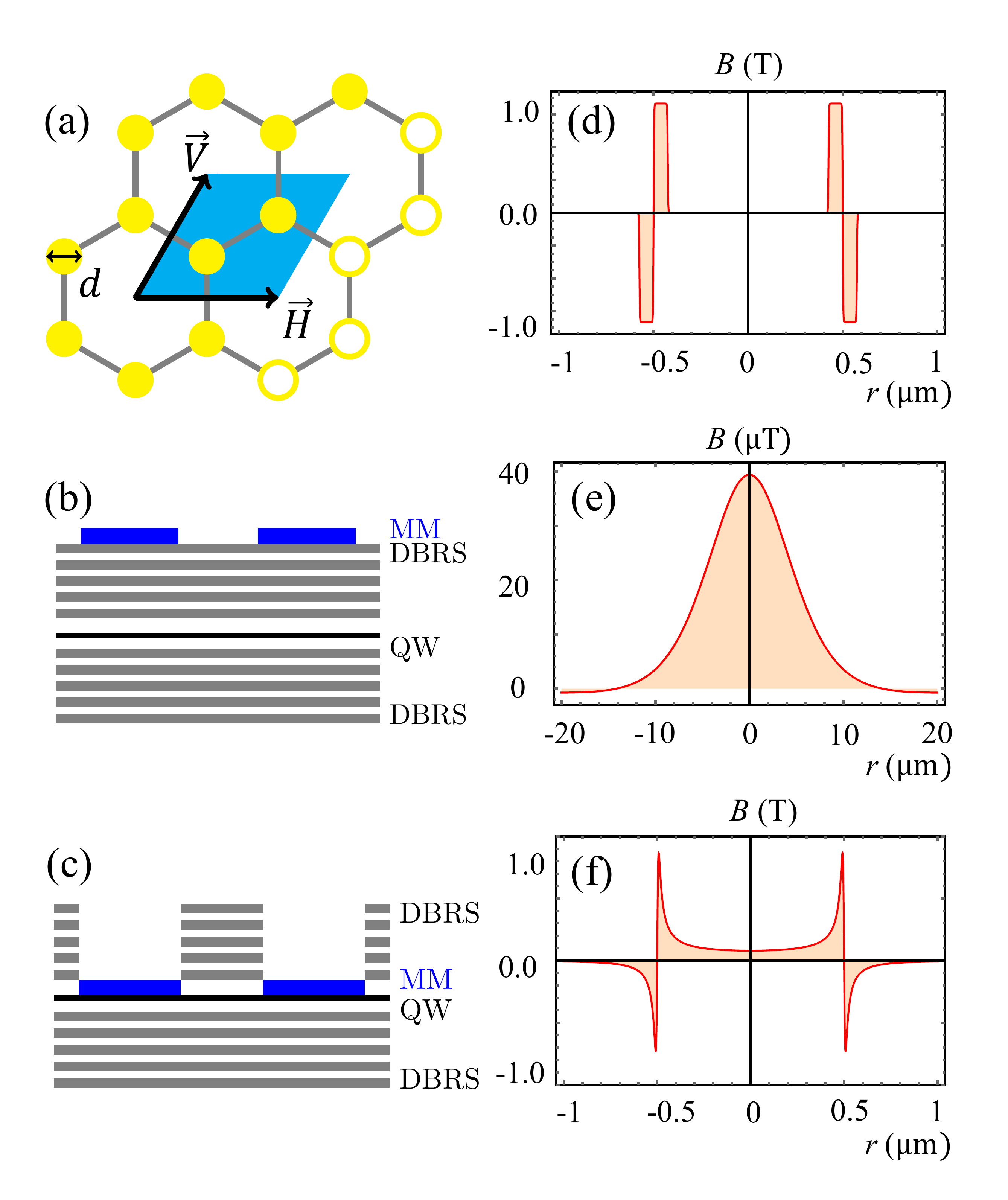}
	\caption{Left panels: System schematic: (a) A honeycomb lattice made of pillars of the etched cavity; yellow empty rings denote the region of the edge. The direction of the edge is along the $\vec{V}$ direction; 
	(b) Setup using magnetic material (MM) placed at the top of the DBR; 
	(c) Setup using MM placed at the top of the QW (close to the exciton layers). 
	Right panels: Magnetic field profile in the case of a single MM disk with radius $R = 0.5$~$\mu$m  (d) Magnetic field used in numerical calculations, rectangle functions with width $dw$; (e) Magnetic field for the case (b) and 10$\mu$m away from the QW; (f) Magnetic field for case (c) and 8~nm away from the QW. }
\label{fig:Fig1}
\end{figure}

Let us consider polaritons loaded in the honeycomb lattice shown in Fig.~\ref{fig:Fig1}(a). 
In recent experiments~\cite{Klembt:2018aa}, an external homogeneous magnetic field perpendicular to the lattice was applied to break time reversal symmetry. 
To get rid of the setup required to generate a sufficiently strong magnetic field, we use a layer of magnetic material (MM), which represents an array of sub-micron magnetic quantum dots. When magnetized they can replace an external field.
%
Let us first consider the case, when the MM is located on the pillars from the top, thus covering the upper Bragg mirror. 
Then the MM represents an array of magnetic quantum dots (QDs) [see Fig.~1(b)].
%
%
%

The magnetic field from a single disc-shaped quantum dot of radius $R$ 
reads~\cite{Erdin:2002aa,Lyuksyutov:2005aa}:
\begin{eqnarray}\label{EqDotMF}
B_z(r,z)=2\pi \mu_0 M R\int_0^\infty J_0(rq)J_1(Rq)\mathrm{e}^{-|z|q}q dq,
\end{eqnarray}
where $\mu_0 = 4\pi\times10^{-7} H/m$ is the vacuum permeability, $M=8\times 10^4Oe\cdot m$ is the two-dimensional magnetization perpendicular to the disc-shaped quantum dot, $J_i$ are the Bessel functions, $r$ is the distance from the center of the QD, $q$ is the in-plane momentum (to be integrated out) and $z$ is the vertical distance from the MM to the QW.

The integral in~\eqref{EqDotMF} can be calculated semi-analytically, using the special functions~\cite{A.-P.-Prudnikov:2003aa}, giving:%
\begin{eqnarray}
B_z=\frac{\left[
f^2(R^2-r^2-z^2)E(f)
+(1-f^2)4rRK(f)
\right]}{4\mu_0^{-1}M^{-1} (Rr)^{3/2}(1-f^2)f^{-1}},~~ \label{Bfield}
\end{eqnarray}
where $f=\sqrt{\frac{4rR}{z^2+(r+R)^2}}$, $K(f)$ and $E(f)$ are the elliptic integrals of the first and second kind, correspondingly. 

Figure~\ref{fig:Fig1}(e) shows the calculated profile of the $z$-projection of the magnetic field. 
Since the distance between the layers of the excitons and the MM amounts to $z\sim$10 $\mu$m in this case (according to a typical size of a microcavity), the resulting effective magnetic field acting on the excitons is only of the order of $\mu$T, which is too small to create a useful topological band gap. This is because $B_z$ in Eq.~\eqref{Bfield} decays exponentially with $z$.

To achieve a strong enough magnetic field, one has to reduce the separation between the MM and QW.
Thus we consider the scheme shown in Fig.~\ref{fig:Fig1}(c), where the microcavity is etched to form a honeycomb lattice and a layer of MM is deposited in the etched regions.
This allows the MM to be placed very close to the quantum wells, where the excitons reside, leading to the interlayer separations of only $z \sim $1--20~nm and strong magnetic fields ($\sim 1$T).
In this case the magnetic field profile [Fig.~\ref{fig:Fig1}(f)] has a very different shape: it is strongly localized at the micropillar/MM boundary, where it also changes sign. 
Thus, the magnetic field is \emph{staggered} with zero net flux.
In Fig.~\ref{fig:Fig1}(f) we show that a MM with the shape of a disk with radius $R=0.5\mu$m produces the magnetic field $\sim 1$~T at the distance 8~nm away.

To prepare the cavity with embedded magnetic quantum dots one should, first, grow the cavity (multilayered structure); second, etch micropillars forming the desired lattice potential; and third, deposit the magnetic material onto the whole structure from the top. This way part of the magnetic material will be on the edges of the pillars. 
However, as our calculations explicitly demonstrate, such remote sources of magnetic field will not give substantial contribution and can be disregarded.
Finally, by heating the sample above the Curie temperature and exposing it to a strong magnetic field, we can magnetize the sample and make it a permanent magnet.


\section{Transport of exciton-polaritons }
To determine whether this kind of staggered magnetic field configuration is capable of inducing topological edge states, we compute the band structure and Chern numbers of the polariton honeycomb lattice numerically. 
Due to the computational limitations (finite discretization), we neglect  weak magnetic field inside and far away from the micropillars, and we approximate $B_z$ with a step function profile with the width $dw$, shown in Fig.~\ref{fig:Fig1}(d).
%
We describe the time evolution of the exciton-polaritons in the cavity by the Gross-Pitaevskii equation,
\begin{eqnarray}
    \label{EOM}
    i\hbar \frac{\partial \psi_{\pm}}{\partial t} &=& -\frac{\hbar^2}{2m_{eff}}\nabla^2 \psi_{\pm} + V \psi_{\pm} + \Delta_{\pm}^{eff} \psi_{\pm} \nonumber \\
    &+& \beta^{eff} \left( \partial_x \mp i \partial_y \right)^2 \psi_\mp, \label{GPE}
\end{eqnarray}
where $\psi_\pm$ are the wave functions of polaritons with up- and down-polarization; $m_{eff}$ is the effective mass; the first term in the r.h.s. stands for the free particle propagation, $V$ is the potential of the honeycomb lattice with the lattice constant $3$~$\mu$m and the diameter of the site $d = 0.75$~$\mu$m; $\Delta^{eff}_\pm$ is the effective Zeeman splitting due to the presence of the MM with the shape presented in Fig.~\ref{fig:Fig1}(d). 
In the calculations we assume the lateral size of MM to be the same as the well of honeycomb lattice and the width of the rectangle functions [Fig.~\ref{fig:Fig1}(d)] $dw = 0.15$~$ \mu$m; $\beta^{eff}$ is an effective TE-TM splitting.
We choose the Hopfield coefficients to be $\abs{X_H}^2 = 1 - \abs{C_H}^2 = 0.3$.
This gives us the effective polariton mass $m_{eff} \approx m_C/\abs{C_H}^2$, with the effective mass of cavity photon $m_C =3.23\cdot10^{-5}m_e$, where $m_e$ is free electron mass.
For polaritons, we have a similar (to excitons) definition of the effective Zeemann splitting and the TE-TM splitting, $\Delta_\pm^{eff} = \abs{X_H}^2 \Delta_\pm$ and $\beta^{eff} = \beta \abs{C_H}^2$. 
In order to quantitatively estimate the magnetic field, we use the Zeeman splitting determined by the relation $\Delta_+-\Delta_- = g_x \mu_B |\mathbf{B}|$, where $g_x \mu_B\approx 180$~$\mu$eV$\cdot$T$^{-1}$ for excitons~\cite{Schneider:2013aa}.
Thus, for a Zeeman splitting term equal $\Delta = \abs{\Delta_\pm}=1$~meV, the peak magnetic strength approximately equals to $5.5$~T which means the distance of MM is $1$~nm away from the QW.

%
%
%
%
%
%


\section{Phase diagram and edge modes} 
First, we compute the bulk bands of Eq.~\eqref{GPE} by assuming a periodic structure and the Bloch wave Ansatz $\psi_{\pm} = u_{\pm}(\mathbf{k},\mathbf{r}) e^{i \mathbf{k}\cdot \mathbf{r} + i E t}$.
Second, we find the bulk Bloch wave eigenstates and the Chern number~\cite{Fukui:2005aa} 
\begin{equation}
C =  \sum_{E_n < E_g} C_n = \frac{1}{2\pi i}\sum_{E_n < E_g} \oint_{\mathrm{BZ}}F_{\mu\nu}^n d^2k,
\label{Chern}
\end{equation}
where the Berry connection $A_n\left(k\right)$ ($n = 1,2$) in the $n$th band below the energy gap ($E_n<E_g$) and the associated field strength $F_{\mu\nu}\left(k\right)$ are defined as
\begin{eqnarray}
    A_\mu &=& \bra{n\left(k\right)}\partial_\mu\ket{n \left(k\right)},\nonumber \\
    F_{\mu\nu}\left(k\right) &=& \partial_\mu A_\nu\left(k\right) - \partial_\nu A_\mu\left(k\right),
\end{eqnarray}
$\ket{n\left(k\right)} $ is the eigenvector of the $n$th Bloch band, and the inner product denotes integration over the unit cell.
Unit vector $\mu$ and $\nu$ denote the directions of the two reciprocal lattice vectors.

\begin{figure}[ht]
    \centering
    \includegraphics[width=0.49\textwidth]{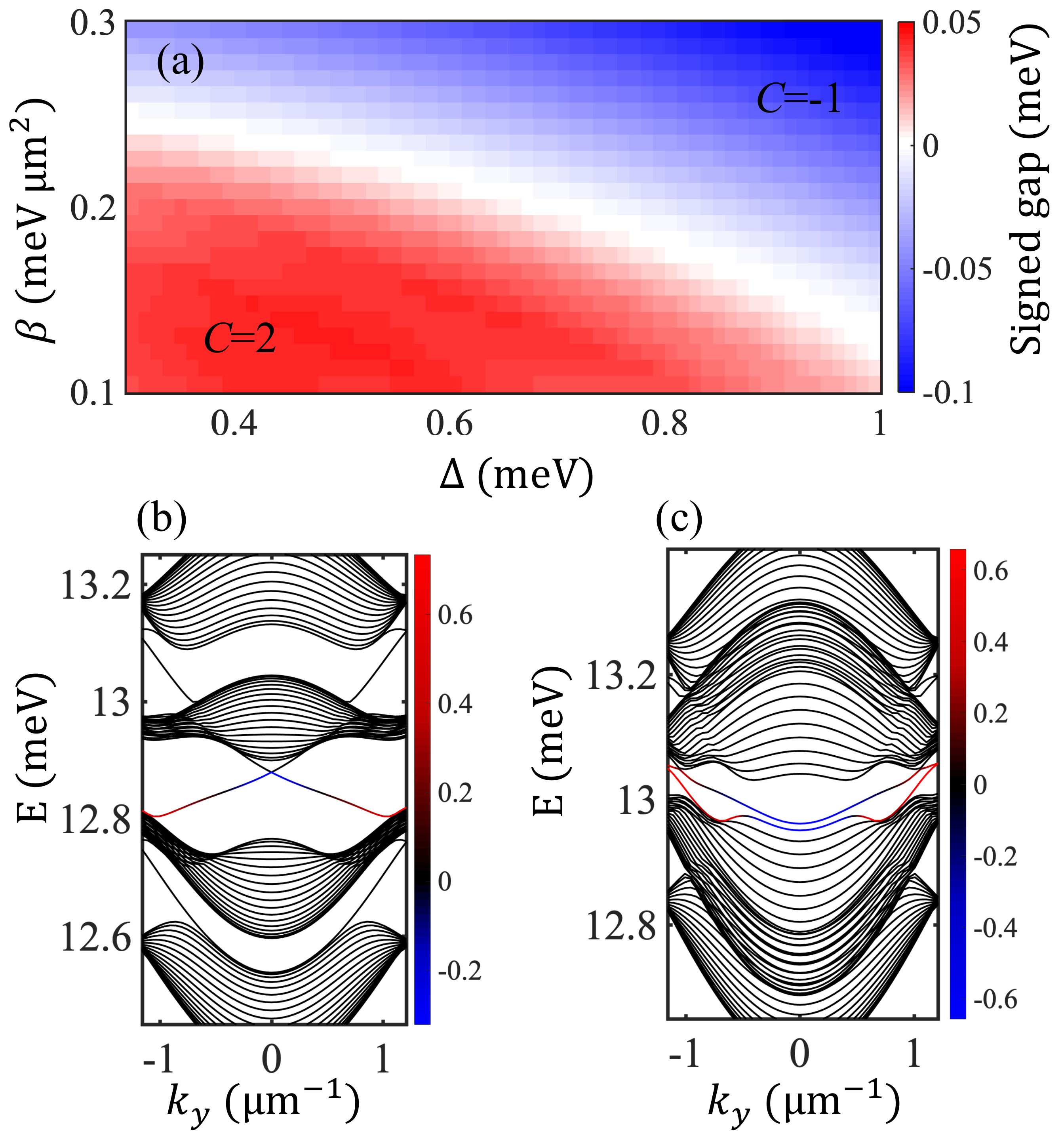}
    \caption{(a) Phase diagram for the Chern numbers with the TE-TM splitting $0.1 \backsim 0.3$~meV$\mu$m$^2$ and Zeeman splitting $0.3 \backsim 1.0$~meV. The color bar indicates the size of the direct gap multiplied by the sign of the Chern number in the units of $\left[meV\right]$ (b) and (c) Spectra of semi-infinite systems with Chern numbers $C = -1$ and $C = 2$, respectively. The color indicates the polarization imbalance, $I = \frac{\abs{\psi_+}^2-\abs{\psi_-}^2}{\abs{\psi_+}^2+\abs{\psi_-}^2}$. For $C=-1$ case, the edge mode is calculated at $\Delta=0.95$~meV and $\beta=0.27$~meV$\cdot\mu$m$^2$. For $C=2$ case, the edge mode was calculated at $\Delta=0.46$~meV and $\beta=0.13$~meV$\mu$m$^2$.}
    \label{fig:Fig2}
\end{figure}

By solving Eq.~\eqref{Chern} for the system described by Eq.~\eqref{EOM}, we can plot the Chern number [Fig.~\ref{fig:Fig2}(a)], using the following parameters: $\Delta_\pm=\left[0.3\sim1.0\right]$~meV and TE-TM splitting is $\beta = \left[ 0.1\sim0.3\right] $~meV$\mu$m$^{2}$.
As shown in Fig.~\ref{fig:Fig2}(a), the system can possess different Chern numbers, depending on the parameters. The phase transition between Chern numbers $C=2$ and $C=-1$ is due to the closing of the gap at the $M$ point and then opening it again. 
A similar behavior has been reported in~\cite{Bleu:2017aa} for the case of a homogeneous external magnetic field.

To verify the existence of topological edge states, we also compute the energy spectrum of a semi-infinite structure (with 20 unit cell in $\vec{H}$ direction and twisted boundary conditions in $\vec{V}$ direction).
Typical spectra for the two topological phases are plotted in Fig.~\ref{fig:Fig2}(b,c), revealing both the bulk bands and band gaps hosting one and two edge states, respectively, in the gap between the second and third bands. 
%
%
%
%
%
%


\section{Comparison with the homogeneous case}
\begin{figure}[!b]
    \centering
    \includegraphics[width=0.40\textwidth]{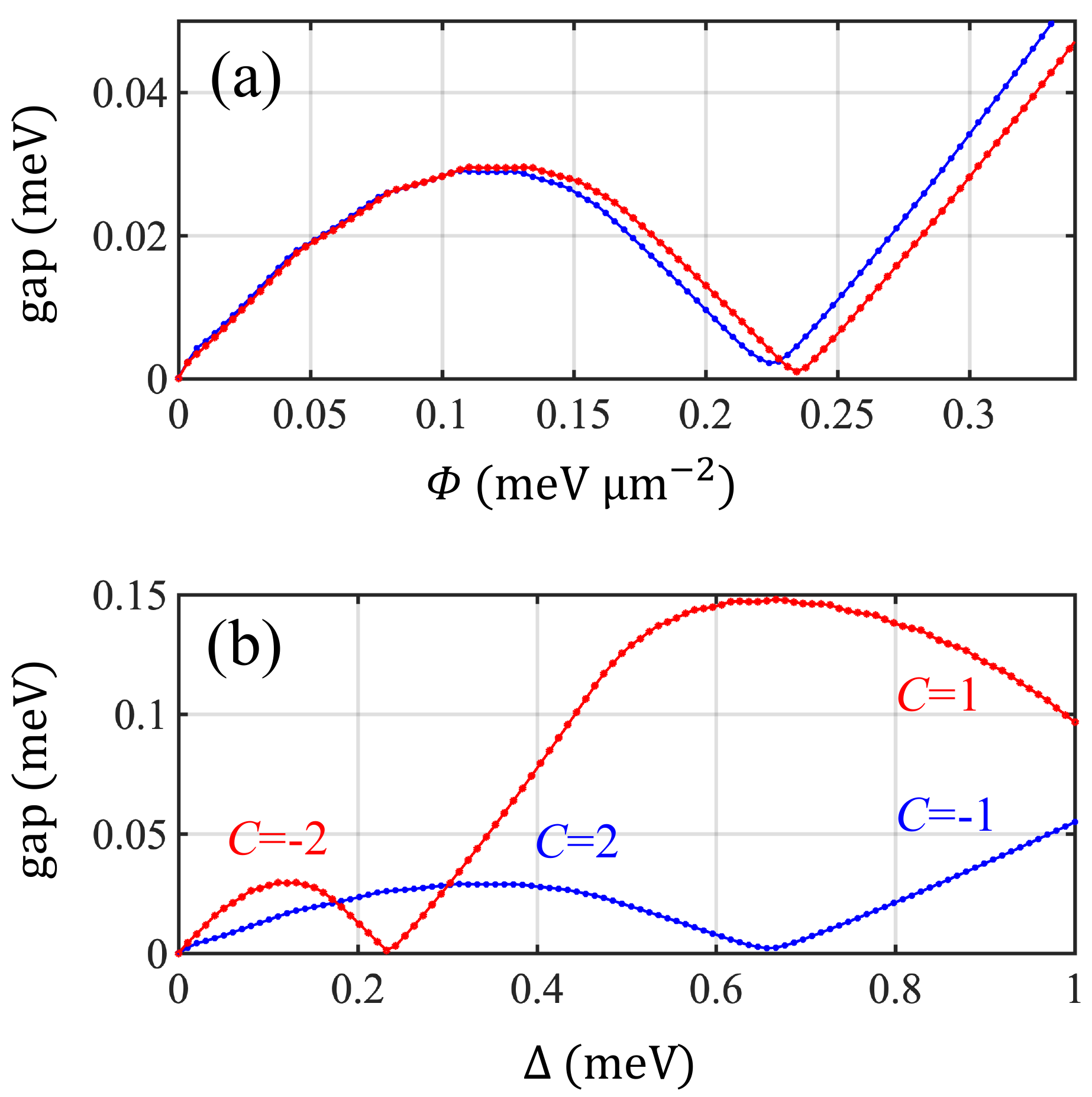}
    \caption{(a) Bandgap as a function of the absolute flux. (b) Bandgap as a function of peak value of the Zeeman splitting. Blue lines show magnetic dot case, red lines a homgeneous magnetic field. The TE-TM splitting is fixed to $\beta = 0.2$~meV$\cdot\mu$m$^2$.}
    \label{fig:gap}
\end{figure}

An important difference between the MM and the external magnetic field cases is the magnitude of the magnetic flux.
In the MM case, the total net flux within one unit cell is almost zero due to strong localization of the magnetic field. 
To compare the MM and the external magnetic field cases, we define the absolute value of the flux on the plane by the integral over a single unit cell region as the unit:
\begin{equation}
    \mathit{\Phi} = \frac{\oint_S \abs{{\Delta_{\pm}}}ds}{S},
\end{equation}
where $S$ is the area of the unit cell. 
Since the Zeeman splitting is proportional to the magnetic field, we can compare the band gaps (between the 2nd and 3rd levels) as functions of the absolute flux in the homogeneous and MM cases.
In what follows we assume the direction of the magnetic field same as the direction of the outer edge of the MM. 
Figure~\ref{fig:gap}(a) shows that given the same value of the absolute magnetic flux, the gap closes and reopens similarly. 

We can also compare the magnitude of the gap as a function of the peak value of the magnetic field, i.e., peak intensity of the Zeeman splitting. 
Figure~\ref{fig:gap}(b) demonstrates the sizes of the gaps in the MM and homogeneous cases as functions of the peak value of the Zeeman splitting. 
We also label the Chern numbers of both the cases before and after the gap is closed.
Obviously, in the homogeneous case, the gap closes much earlier than in the MM case. 
%
%
%


\section{Cavity with a MM exposed to an external magnetic field}
\begin{figure}[hb]
\centering
\includegraphics[width=0.48\textwidth]{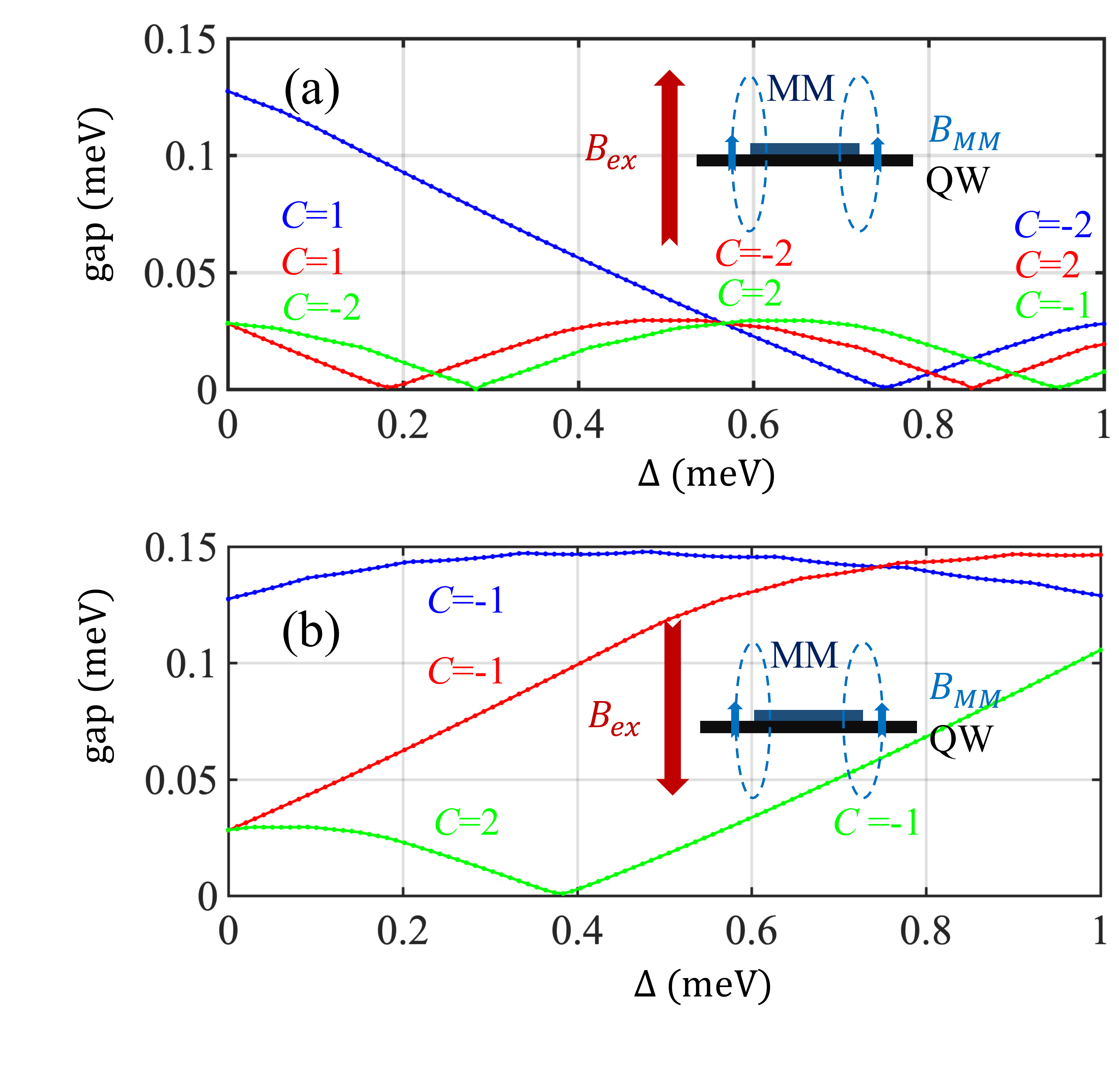}
\caption{The mutual effect of MM and external magnetic field. 
The x-axis labels the intensity of the local magnetic field of MM. 
Different colors reflect different magnitudes of external magnetic field, 
$B_{ex}\approx 2.8$, $1.7$, and $0.5$~T, corresponding to the Zeeman splittings $\Delta_{ex}=0.5$ (blue), $0.3$ (red), and $0.1$~meV (green).
The Chern numbers are labeled before and after each gap.
The TE-TM splitting is fixed at $\beta = 0.2$~meV$\cdot\mu$m$^2$.}
    \label{fig:fig4}
\end{figure}

Having a cavity with embedded MM, one can in addition apply a homogeneous magnetic field to the sample. The resulting field represents a superposition of two magnetic field profiles (Fig.~\ref{fig:fig4}).

Let us first consider the case when the external magnetic field $\mathbf{B}_{ex}$ is parallel to the outside edge of the MM, $\mathbf{B}_{MM}$, as shown in inset of Fig.~\ref{fig:fig4}(a).
In this situation, we see further additional phase transition in red and green curves.
Specifically, we can classify two distinct regimes.
The first one is when the MM is weak ($\Delta < 0.2$~meV). With the increasing homogeneous magnetic field ($\Delta_{ex}$, from green to red curve), the system undergoes the phase transition from Chern number $C=-2$ to $C=1$;
Second, when the MM is moderately strong ($0.3$~meV$<\Delta<0.8$~meV), the Chern number changes from $C=2$ to $C=-2$ as $\Delta_{ex}$ increases.

Figure~\ref{fig:fig4}(b) shows the effect of the oppositely-directed external magnetic field. 
Red and blue curves show that we can efficiently enlarge the size of the gap in the $C=-1$ case. 
The size of the gap is vital in topological polaritonic systems since the finite lifetime of the exciton-polaritons can broaden the bandwidth of the spectrum and diminish and destroy nontrivial topological properties.
We also conclude that the combination of two magnetic fields  reduces the strength of the external magnetic field required to generate a topological insulator state with a given band gap.
%
%
%



\section{Discussion and conclusions}
We have shown that a local magnetic field due to the presence of a magnetic material can be sufficiently strong to open a gap at the Dirac point and allow for the observation of nontrivial topological states in an exciton-polariton system loaded in a honeycomb lattice. 
With the change of the intensity of the built-in magnetic field or the TE-TM splitting, the system undergoes a phase transition between two nontrivial states with the Chern numbers $\pm2$ and $\mp1$.

The key advantage of this setup is the size of the system, which can be much smaller than the one requiring a homogeneous external magnetic field. 
This can be highly beneficial for future experiments and applications in devices. 
Furthermore we have studied the Chern numbers and gap sizes as functions of the magnetic flux strength and the peak value of the magnetic field.
The results show that the design with the use of magnetic material and the regular homogeneous case demonstrate similar behavior.

Furthermore, we have explored the joint effect of the internal magnetic material field with an external magnetic field. 
Depending on the relative direction of the two fields, one can switch between the Chern numbers. 
This switching can be performed ``on the fly", since it is only dependent on a small change of external magnetic field, thereby enabling control over the number and/or direction of the topological edge states.
By reversing the direction of the external magnetic field, one can also keep the Chern number the same but enlarge the size of the gap significantly, thus increasing the speed of the edge state. 
This allows us to propagate polaritons over longer distances before they decay due to their finite lifetime.

\section*{Acknowledgements}
We thank Prof. D.~Solnyshkov for useful discussions. This work was supported by the Institute for Basic Science in Korea (Projects IBS-R024-D1 and IBS-R024-Y1).

\bibliography{lib}

\end{document}